**Distribution fitting 12. Sampling distribution of compounds abundance from plant species measured by instrumentation. Application to plants metabolism classification**

**Lorentz JÄNTSCHI[1,3], Sorana D. BOLBOACĂ[2], Radu E. SESTRAŞ[1]**

[1] University of Agricultural Sciences and Veterinary Medicine Cluj-Napoca
[2] University of Medicine and Pharmacy "Iuliu Haţieganu" Cluj-Napoca
[3] Technical University of Cluj-Napoca
[*] Corresponding author: lori@academicdirect.org

**Abstract.** A series of ten plant species belonging to *Magnoliopsida - Dicotyledons* class were analyzed in terms of chemical compounds distribution of abundance, starting from the assumption that these distributions should give a picture of similarities and differences between plants metabolism. From a pool of theoretical distributions, log-normal distribution was selected giving the best accuracy with the modeled phenomena and agreement with the observed data. From obtained lognormal distributions statistics a classification were constructed and were compared with the classification based on phylogeny.

### Introduction

There are many studies regarding the plants classification and phylogeny, starting basically from two approaches: the morphology based one (Grime & others, 1997), and from molecular based one (Sonnhammer & others, 1997).

Analysis of statistical distributions may provide useful evidences for plants evolution. According to (Fay & Wu, 2000), frequency distribution of variation can be influenced by a number of evolutionary processes, an excess of derived variants at high frequency is a unique pattern produced by hitchhiking.

The distribution analysis extends their application in recent years, from (Reed & others, 1985) to ecological modeling of natural structures assemblages (Tsirtsis & Spatharis, 2011) and morphologic parts modeling (Alarcon & Sassenrath, 2011).

The present study propose another approach for plants classification and kinship, based on similarities on distribution of chemical compounds, which should be related to plants metabolism.

### Material

Data reported in (Kozioł & Macía, 1998) and (Soković & others, 2007) were included into analysis. Table 1 contains the values of compounds abundances (as were reported) sorted descending for ten different species.

Tab 1.
Compounds abundances (percent values) for ten different species

| Series | Compounds abundances (count) USDA Label (*Species name*) [Ref] |
|---|---|
| Ob% | 69.25; 2.56; 2.48; 2.38; 2.10; 1.87; 1.66; 1.42; 1.13; 1.11; 1.05; 1.02; 0.91; 0.82; 0.82; 0.63; 0.58; 0.56; 0.51; 0.46; 0.43; 0.43; 0.39; 0.38; 0.31; 0.30; 0.30; 0.27; 0.19; 0.12; 0.11; 0.10; 0.09; 0.06; 0.06; 0.05 (36) OCBA (*Ocimum basilicum*) [8] |
| Sp% | 38.99; 18.51; 6.95; 3.07; 1.30; 1.29; 1.19; 0.89; 0.88; 0.76; 0.66; 0.55; 0.52; 0.49; 0.47; 0.45; 0.40; 0.26; 0.22; 0.21; 0.20; 0.19; 0.12; 0.11; 0.08; 0.07; 0.05; 0.04; 0.03 (29) SPPU (*Spondias purpurea*) [7] |
| Mp% | 37.40; 17.37; 12.70; 6.85; 6.82; 5.59; 2.52; 1.29; 1.23; 0.81; 0.79; 0.69; 0.50; 0.48; 0.47; 0.41; 0.29; 0.28; 0.19; 0.17; 0.17; 0.13; 0.13; 0.12; 0.10; 0.10 (26) MEPI (*Mentha piperita*) [8] |
| Ms% | 49.52; 21.92; 5.77; 3.06; 2.28; 1.36; 1.27; 0.99; 0.71; 0.71; 0.68; 0.57; 0.52; 0.49; 0.49; 0.48; 0.45; 0.40; 0.33; 0.31; 0.30; 0.26; 0.22; 0.07 (24) MESP3 (*Mentha spicata*) [8] |
| Tv% | 48.92; 18.99; 4.08; 3.45; 3.45; 2.23; 1.78; 1.73; 1.72; 1.30; 1.21; 1.17; 1.06; 0.83; 0.76; 0.74; 0.65; 0.58; 0.46; 0.41; 0.33; 0.30; 0.17; 0.16 (24) THVU (*Thymus vulgaris*) [8] |
| So% | 31.65; 16.67; 8.70; 6.90; 4.77; 4.61; 3.41; 3.03; 2.64; 2.56; 2.20; 1.74; 1.09; 0.99; 0.37; 0.35; 0.30; 0.29; 0.14; 0.12; 0.11; 0.07; 0.03 (23) SAOF2 (*Salvia officinalis*) [8] |
| Mc% | 43.47; 9.09; 8.50; 8.48; 6.06; 5.62; 5.21; 1.92; 1.65; 0.39; 0.38; 0.35; 0.35; 0.32; 0.29; 0.16; 0.15; 0.12; 0.10; 0.08 (20) MARE6 (*Matricaria chamomilla*) [8] |
| La% | 27.54; 27.21; 8.50; 6.54; 4.20; 3.34; 2.95; 2.51; 2.44; 2.09; 2.02; 1.07; 0.59; 0.58; 0.25; 0.19; 0.16; 0.09; 0.06; 0.04 (20) LAAN81 (*Lavandula angustifolia*) [8] |
| Cl% | 59.68; 17.25; 11.21; 2.85; 1.72; 1.29; 0.87; 0.84; 0.64; 0.55; 0.44; 0.39; 0.29; 0.27; 0.21; 0.17; 0.13 (17) CILI5 (*Citrus limon*) [8] |
| Ov% | 64.50; 10.90; 10.80; 3.50; 2.50; 2.20; 2.20; 1.90 (8) ORVU (*Origanum vulgare*) [8] |
| Refs: | [7] - (Kozioł & Macía, 1998); [8] - (Soković & others, 2007) |

**Methods**

It should be noticed that always the sum of percentage values given in Table 1 does not exceed 100%; more than that, never the sum of values does not give 100%, always being something undetermined (unknown compounds or compounds with abundance below the detection limit of the instrumentation). Thus the assumption of independence between the percentage values is sustained.

Nerveless, it may be dependence between the occurrences or abundance of certain compounds, given by the intrinsic biological processes through these compounds was synthesized in plants. In fact, the main issue addressed with our analysis addresses exactly this intrinsic dependence.

Let us assume that we will observe the river flow frequency distribution (by defining a constant time step interval). Then we may observe that flows follow a log-Pearson Type 3 distribution (Khan, 2009). We may found also that log-Pearson Type 3 distribution is the guideline recommended distribution when we deal with water flows (US Water Resources Council, 1967). But these values of flood flow are indeed independent one to each other? Or after a period when flood flow are high, are expected a long period with low flood flows, and this is the main reason for which log-Pearson Type 3 distribution fits very well? This is a very good example on how the intrinsic relationship between observable values is propagated to the distribution type. Same reasoning stays at the basis of discovery and usage of any other distribution, that the distribution extracts the mimic existing in our sampled data and provides the shape of the population based on this mimic. For example, when different people made same measurements of same observable in same place at same moment, regardless the type of the observable and regardless of the type of the instrumentation involved, the only one assumption which we can made is that their measurements should follow a symmetrical distribution, because is equiprobable to make an error observing less or observing more. If we know more about the measurement procedure, such as were made on a absolute or relative scale, were made using an instrumentation or only a hand tool, observation results were recorded on a continuous (or close to continuous) or a discrete scale, on a ordinal or category based scale), are a finite or a infinite number of possible observations, so on, then we are able to narrow the pool of possible theoretical distributions of the observable results to a short list of probable distributions.

Regarding this last statement, that knowing more about the type of measurement involved we may narrow the pool of possible distributions to a short list of probable distributions, this are based on evidences too. There are two types of evidences which may come: from theoretical considerations of the phenomena involved, when we arise again to our previous example of flood flow frequencies, or from experimental considerations, when a great number of independent observations suggesting same conclusion. In one way or another, when we arisen to a conclusion we transfer the knowledge from applications to theory or vice versa.

The data from Table 1 were obtained in one case only (one sampling) from the entire population of species individuals and the true values of the abundances may be different in samples than in population from which the samples were drawn. More than that, for every analyzed compound, its abundance in population is not identical from one individual to another, and it has variability, and it follows a distribution characterized by a mean (true value of the abundance in population) and a standard deviation. The idea is to explore the distribution of the abundances of the compounds in these different species (observed distributions, and then the distributions will be sampling distributions of abundances) and to extract useful information which may be true in general, for every chemical analysis of any species.

### Results

A pool of over 50 theoretical distributions were narrowed to a short list of 16 possible distributions for the populations of chemical compounds abundances in species from Table 1, based on "have fit with observed values" and having probability to come from the theoretical distribution greater than 1% when one of Kolmogorov-Smirnov (K-S) - (Kolmogorov, 1941; Smirnov, 1948), Anderson-Darling (A-D) - (Anderson & Darling, 1952), and Chi-Square (C-S) - (Pearson, 1900; Fisher, 1922a; Fisher, 1924) statistics were applied to measure the agreement between observations and theoretical distributions of which parameters were determined using Maximum Likelihood Estimation (MLE) - (Fisher, 1912; Fisher, 1925) method.

These distributions are: Dagum (3 Parameters), Frechet (2P), Frechet (3P), Fisher-Tippett (3P), Inverse Gaussian (2P), Inverse Gaussian (3P), Levy (1P), Levy (2P), Log-Logistic (2P), Lognormal (2P), Pareto 2, Pearson 5 (2P), Pearson 5 (3P), Pearson 6 (3P), Phased Bi-Exponential (4P), Weibull (2P). K-S, A-D and C-S statistics and associated probabilities (that samples come from the theoretical distributions) are given in Table 2. Since every statistics measures the departure between the observation and the model using different approaches, a global measure of likelihood were calculated for a given Probability Density Function expression by using the formula proposed in (Fisher, 1948) - F-C-S column in Table 2.

Two distributions were removed from further analysis at this step (Dagum - 3P and Inverse Gaussian - 2P - see Table 2) due to their unacceptable probability of observation ($p_{F-C-S}$ column in Table 2).

Tab. 2

Measuring agreement between observations and theoretical distributions

| Dist | C-S | $p_{C-S}$ | K-S | $p_{K-S}$ | A-D | $p_{A-D}$ | F-C-S | $p_{F-C-S}$ | ΣCDF0 | ΣCDF100 | Rank |
|---|---|---|---|---|---|---|---|---|---|---|---|
| Dagum_3P | 16.48 | 0.0574 | 17.74 | 0.0595 | 28.11 | 0.0017 | 12.04 | 0.0073* | *rejected at this stage | | 15 |
| Frechet_2P | 5.51 | 0.7874 | 6.27 | 0.7918 | 7.29 | 0.6974 | 0.83 | 0.8416 | 0 | 0.183 | 13 |
| Frechet_3P | 2.73 | 0.9742 | 3.17 | 0.9771 | 4.59 | 0.9166 | 0.14 | 0.9871 | 0.0018 | 0.302 | 5 |
| FisherTippett_3P | 6.80 | 0.6576 | 7.33 | 0.6941 | 9.53 | 0.4831 | 1.51 | 0.6795 | 0.7922 | 0.035 | 10 |
| InverseGaussian_2P | 28.50 | 0.0008 | 27.92 | 0.0019 | 44.26 | 0.0000 | 26.17 | 0.0000* | *rejected at this stage | | 16 |
| InverseGaussian_3P | 3.05 | 0.9625 | 7.58 | 0.6694 | 7.41 | 0.6863 | 0.82 | 0.8456 | 0.0117 | 0.061 | 12 |
| Levy_1P | 11.89 | 0.2196 | 14.28 | 0.1606 | 12.26 | 0.2679 | 4.66 | 0.1983 | 0 | 0.562 | 7 |
| Levy_2P | 4.38 | 0.8849 | 8.86 | 0.5458 | 8.26 | 0.6031 | 1.23 | 0.7450 | 0.0033 | 0.438 | 11 |
| LogLogistic_2P | 2.69 | 0.9754 | 4.60 | 0.9162 | 6.72 | 0.7514 | 0.40 | 0.9406 | 0 | 0.085 | 2 |
| Lognormal_2P | 5.50 | 0.7887 | 6.88 | 0.7364 | 7.69 | 0.6589 | 0.96 | 0.8108 | 0 | 0.033 | 1 |
| Pareto2_2P | 6.07 | 0.7325 | 5.14 | 0.8816 | 7.40 | 0.6869 | 0.81 | 0.8464 | 0 | 0.108 | 3 |
| Pearson5_2P | 4.81 | 0.8504 | 6.17 | 0.8004 | 6.97 | 0.7279 | 0.70 | 0.8726 | 0 | 0.237 | 9 |
| Pearson 5_3P | 3.37 | 0.9479 | 5.11 | 0.8837 | 17.53 | 0.0634 | 2.94 | 0.4016 | 0.0015 | 0.373 | 6 |
| Pearson6_3P | 3.93 | 0.9158 | 2.88 | 0.9841 | 4.67 | 0.9118 | 0.20 | 0.9782 | 0 | 0.111 | 4 |
| PhasedBiExponential_4P | 3.44 | 0.9442 | 6.30 | 0.7895 | 17.20 | 0.0701 | 2.95 | 0.3993 | 0.0000 | 0.060 | 8 |
| Weibull_2P | 13.21 | 0.1533 | 6.10 | 0.8070 | 18.02 | 0.0546 | 5.00 | 0.1720* | *rejected at this stage | | 14 |

The data comes from measurements of percentages, and thus may vary from 0(%) to 100(%) and is essential that the theoretical distributions to reflect this fact as best as possible (if is possible via domain of the theoretical distribution, or at least based on cumulative probabilities in these critical points). Addressing this issue, cumulative probabilities for X≤0 (ΣCDF0 in Table 2, as sums from all 10 samples) and for X≥100 (ΣCDF100 in Table 2, as sums from all 10 samples) were calculated. When the distribution domain is strictly positive by definition a value of "0" were reported in CDF0 column of Table 2, or "0.00" or greater otherwise. The distributions were ranked (Rank column in Table 2) on ΣCDF0 as first criterion (ascending order, with "0.000" > "0"), ΣCDF100 as second criterion (ascending order, "0.000" > "0" were not necessary to be applied), and $p_{F-C-S}$ (descending order - were not necessary to be applied).

It should be noticed that any of these criteria are not absolute and nor the ranking is. Nevertheless, it should be noticed too that the ranks from $p_{F-C-S}$ are reversed for first two pole positions (Log-Logistic and Log-Normal). This is an expected result. Both distributions compete for same pool of observed data, as other authors also recently observed (Dey & Kundu, 2010[20]).

More, log-logistic distribution is more tailed than the log-normal. With a kurtosis of 4.2, the standard logistic distribution has a longer tail than the normal, which has kurtosis 3.0; differences in the upper quantiles of normal and logistic are further magnified when they are exponentiated to get log-normal and log-logistic distributions (Modarres & others, 2002). Our results shown this aspect (ΣCDF100 is 0.033 for log-normal and 0.085 for log-logistic under disfavor of log-logistic distribution) - we don't want a distribution to predict much outside of the true domain of the possible data.

Once the log-normal distribution were selected from the pool of the possible distributions, a good idea is to check if the joined pool of data from all species may come from same log-normal distribution. It is a reasoning to check this. Because all species also come from same or different genus, same or different family, same or different order (so on), we may and must assume that a certain level of phylogeny our conclusion regarding the distribution of compounds abundances should be verified (remaining the same).

This hypothesis were verified and proved to be true. The pool of 227 observations (considered drawn from *Magnoliopsida - Dicotyledons* class of *Magnoliophyta* - Flowering plants division). Joined pool of 227 observations has a 23.5% probability to be drawn from lognormal distribution according to K-S statistic, 20.4% probability according to A-D, 31.9% probability according to C-S, a value of 4.18 for F-C-S, and a probability of 24.3% to be drawn from lognormal distribution according to Fisher's method of combining independent tests of significance (calculated as probability from C-S distribution to observe -ln(0.235)-ln(0.204)-ln(0.319) with three degrees of freedom).

Once again, is no reason to reject the hypothesis of lognormal distribution of the data of compounds abundances sampled from biological organisms!

The parameters of the obtained distributions are given in Table 3.

Tab.3
Distributions of the populations of compounds abundances

| Label | Lognormal distribution parameters | Mean | StDev | lnSk | lnKE | FI | H.5 | H1 | H2 | H3 |
|---|---|---|---|---|---|---|---|---|---|---|
| ORVU | Lognormal(x;1.1515; 1.6653) | 10.3 | 17.1 | 2.26 | 5.84 | .754 | 3.37 | 3.23 | 2.74 | 2.59 |
| CILI5 | Lognormal(x;1.6676; 0.00927) | 4.05 | 15.8 | 4.26 | 11.3 | .360 | 3.02 | 1.94 | 1.09 | 0.79 |
| LAAN81 | Lognormal(x;1.842; -0.02378) | 5.33 | 28.6 | 5.14 | 13.6 | .295 | 3.13 | 2.01 | 1.00 | 0.65 |
| MARE6 | Lognormal(x;1.8645; 0.1644) | 6.70 | 37.5 | 5.26 | 14.0 | .288 | 3.25 | 2.21 | 1.18 | 0.82 |
| SAOF2 | Lognormal(x;1.8178; 0.09447) | 5.74 | 29.4 | 5.01 | 13.3 | .303 | 3.19 | 2.11 | 1.13 | 0.78 |
| THVU | Lognormal(x;1.4238; -0.16437) | 2.34 | 6.00 | 3.20 | 8.38 | .493 | 2.63 | 1.61 | 0.95 | 0.71 |
| MESP3 | Lognormal(x;1.3062; 0.18882) | 2.84 | 6.02 | 2.77 | 7.20 | .586 | 2.77 | 1.87 | 1.30 | 1.08 |
| MEPI | Lognormal(x;1.7061; -0.24783) | 3.35 | 13.9 | 4.44 | 11.8 | .344 | 2.88 | 1.71 | 0.82 | 0.51 |
| SPPU | Lognormal(x;1.6655; -0.75731) | 1.88 | 7.27 | 4.25 | 11.2 | .361 | 2.46 | 1.17 | 0.33 | 0.02 |
| OCBA | Lognormal(x;1.3718; -0.60725) | 1.40 | 3.29 | 3.01 | 7.84 | .531 | 2.17 | 1.13 | 0.50 | 0.28 |
| Magnoliopsida | Lognormal(x;1.6744; -0.13751) | 3.54 | 13.9 | 4.29 | 11.3 | .357 | 2.93 | 1.80 | 0.94 | 0.64 |
| Legend: LnSk - ln(Skewness); lnKE - ln(Kurtosis Excess); FI - Fisher's information; Hα - Renyi's Entropies | | | | | | | | | | |

**Discussion**

Figure 1 depicts in logarithmic scale on both axes the log-normal distributions of compounds abundances for investigated species, and Figure 2 depicts their classification.

The Figure 1 were obtained by using the classification data from Cronquist system (Cronquist, 1981) by using different encodings for different values of classifiers. On the tabulated data for the 10 samples of species were applied the cluster analysis method using single linkage based on Euclidian distances. By using the same classification method as were used to obtain Figure 2 by using now the results given in Table 3 for chemical compounds abundance distributions, another classification were obtained, and is given in Figure 3.

Figure 1 shows a wide variety of abundances. Practically every species seems to be specialized in its own way in synthesizing chemicals. Let us note that even if we can see very good associations in terms of distribution of chemical compounds in Figure 1 (as between SAOF2 - *Salvia officinalis* and SPPU - *Spondias purpurea*) there are great differences between these (for example SAOF2 and SPPU belongs to different subclasses - see Figure 2). It is a simple reasoning too seen this: distribution of the compounds abundances give only one component from the whole picture of relatives; another component is for example the structures or the belonging classes of compounds synthesized - and this component should be somehow orthogonal on the component of compounds distribution and giving thus a completely different picture of relatives.

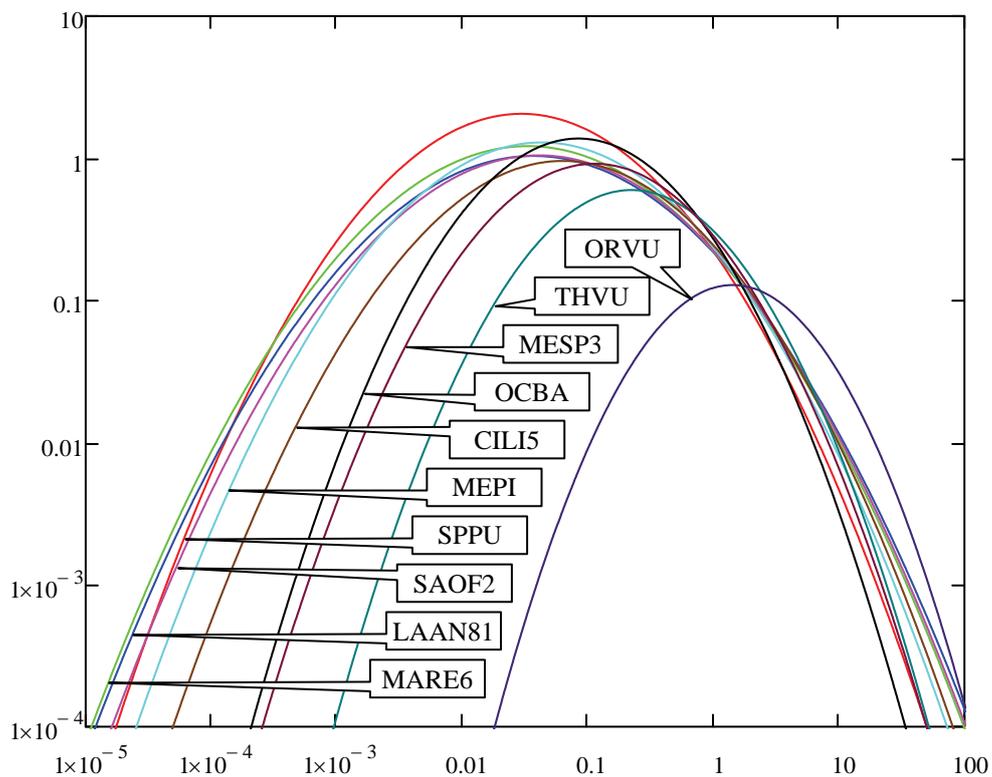

Fig. 1. Chemical compounds abundances for ten species

Legend: MARE6 - *Matricaria chamomilla*; LAAN81 - *Lavandula angustifolia*; SAOF2 - *Salvia officinalis*; SPPU - *Spondias purpurea*; MEPI - *Mentha piperita*; CILI5 - *Citrus limon*; OCBA - *Ocimum basilicum*; MESP3 - *Mentha spicata*; THVU - *Thymus vulgaris*; ORVU - *Origanum vulgare*

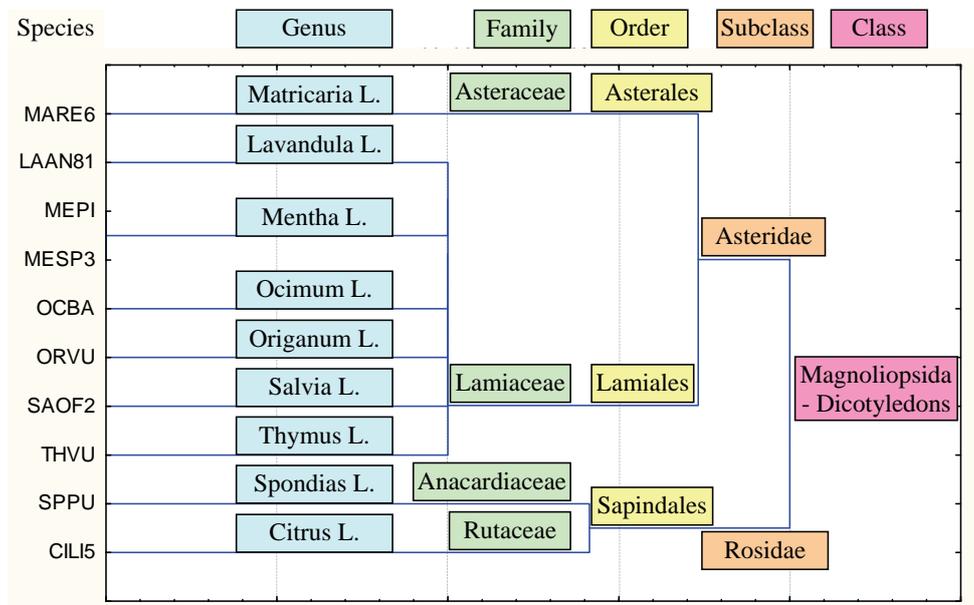

Fig. 2. Investigated species - classification using taxonomic characters of phylogenetics importance

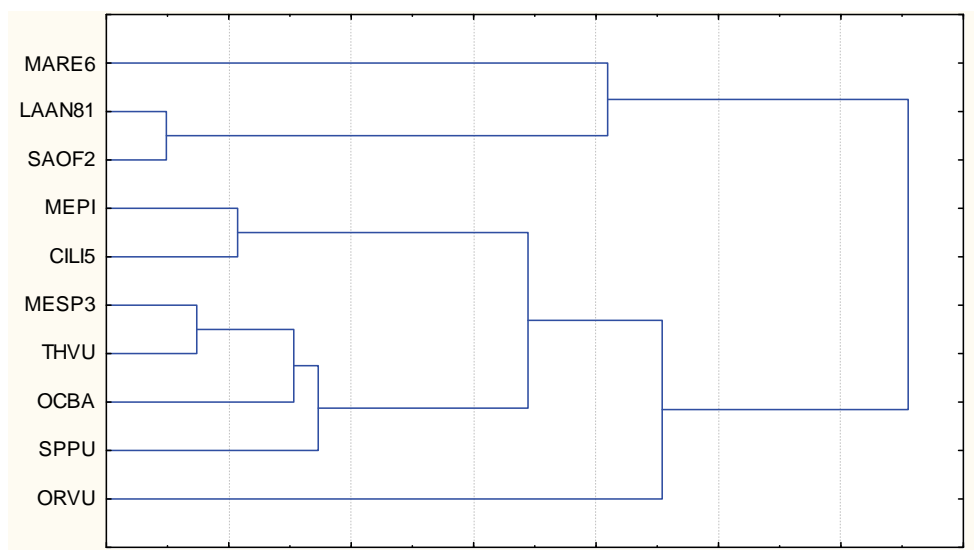

Fig. 3. Investigated species - classification using chemicals abundances distribution statistics

Even further, the classification based on taxonomic characters of phylogenetics importance and the classification based on chemical compounds abundances distribution has no obvious association relationship (as can be seen from the two classifications depicted in Figure 2 and Figure 3). The main reasoning for this should be found in the meaning behind of the data from which the classification was made. Thus, if taxonomic characters give a picture about phylogeny, on the opposite, compounds distribution give a picture about metabolism.

**Conclusions**

Distribution of chemical compounds was used to classify 10 species according to their relatives on the relative ratios of synthesized chemical compounds.

The study showed that with a very high likelihood distribution of chemical compounds in plant species follow a log-normal distribution. Distribution remains the same if in place of a species are placed a plant class, and this fact suggests that splitting of the plants into classes, subclasses, orders, families, genus and species is consistent with plant metabolism too.

Anyway, classification based on distribution of chemical components give a totally different picture of relatives than the phylogeny based classification, suggesting that the classification based on distribution of chemical components is only one component (of great importance for plant species characterization) - the one relating the plant metabolism - from the whole pool of components which gives relatives based on phylogeny, and another component of great importance for plant species characterization should be constructed from chemical compounds similarities.

**Acknowledgments**

The study was supported by POSDRU/89/1.5/S/62371 through a postdoctoral fellowship for L. Jäntschi.